\documentclass[10pt]{article}
\usepackage[utf8]{inputenc}
\usepackage{amsfonts}
\usepackage{amsmath,bm,amsthm}
\usepackage{float}
\usepackage{bbold}
\usepackage{graphicx,rotating,pdflscape}
\usepackage[dvipsnames]{xcolor}
\usepackage{dsfont}
\usepackage[numbers]{natbib}
\usepackage{hyperref}
\usepackage{comment}
\usepackage[normalem]{ulem}
\usepackage{changepage}

\usepackage{tablefootnote,array,multirow}
\usepackage{makecell}
\newcolumntype{L}[1]{>{\raggedright\let\newline\\\arraybackslash\hspace{0pt}}m{#1}}
\newcolumntype{C}[1]{>{\centering\let\newline\\\arraybackslash\hspace{0pt}}m{#1}}
\newcolumntype{R}[1]{>{\raggedleft\let\newline\\\arraybackslash\hspace{0pt}}m{#1}}
\newcolumntype{P}[1]{>{\centering\arraybackslash}p{#1}}
\newcolumntype{M}[1]{>{\centering\arraybackslash}m{#1}}
\newcommand{\titlestudy}[1]{\citeauthor{#1} (\citeyear{#1})}
\usepackage{tikz}
\def\checkmark{\tikz\fill[scale=0.4](0,.35) -- (.25,0) -- (1,.7) -- (.25,.15) -- cycle;}
\usepackage[margin=1.5in]{geometry}
\usepackage{placeins}

\usepackage{forest}
\usetikzlibrary{fit,positioning}
\usetikzlibrary{calc,backgrounds}
\colorlet{pathcolor}{BurntOrange}
\colorlet{refcolor}{Blue}
\tikzset{
  path arrow/.style={
   midway,pathcolor,sloped,fill, minimum height=2.5cm, single arrow, single arrow head extend=.5cm, single arrow head indent=.25cm,xscale=0.3,yscale=0.15,
    allow upside down
  },
  black arrow/.style 2 args={-stealth, shorten >=#1, shorten <=#2},
  black arrow/.default={1mm}{1mm},
  tree box/.style={draw, rounded corners, inner sep=0.5em,minimum width = 15em,minimum height = 10em,outer xsep=0em},
  node box/.style={white, draw=black, text=black, rectangle, rounded corners},
}
\newlength{\arrowdist}
\title{A new methodology to predict the oncotype scores based on clinico-pathological data with similar tumor profiles}
\author{
Zeina Al Masry$^1$, Romain Pic$^2$, Clément Dombry$^2$, Christine Devalland$^3$
}
\date{%
\small
    $^1$Institut FEMTO-ST, Université Bourgogne Franche-Comté, CNRS, SUPMICROTECH-ENSMM, 25 rue Savary, Besançon, France\\ %
   $^2$Laboratoire de Mathématiques de Besançon, CNRS UMR 6623, Univ. Bourgogne Franche-Comté, Besançon, France\\%
   $^3$Service d’anatomie et cytologie pathologiques, Hôpital Nord Franche-Comté, 100 Route de Moval, 90400 Trévenans, France. \\%
}

\begin{document}

\maketitle

\begin{abstract}
\noindent
\textbf{Introduction:} The Oncotype DX (ODX) test is a commercially available molecular test for breast cancer assay that provides prognostic and predictive breast cancer recurrence information for hormone positive, HER2-negative patients. The aim of this study is to propose a novel methodology to assist physicians in their decision-making. 

\noindent
\textbf{Methods:} A retrospective study between 2012 and 2020 with 333 cases that underwent an ODX assay from three hospitals in Bourgogne Franche-Comté was conducted. Clinical and pathological reports were used to collect the data. A methodology based on distributional random forest was developed using 9 clinico-pathological characteristics. This methodology can be used particularly to identify the patients of the training cohort that share similarities with the new patient and to predict an estimate of the distribution of the ODX score. 

\noindent
\textbf{Results:} The mean age of participants id 56.9 years old. We have correctly classified $92\%$ of patients in low risk and $40.2\%$ of patients in high risk. The overall accuracy is $79.3\%$. The proportion of low risk correct predicted value (PPV) is $82\%$. The percentage of high risk correct predicted value (NPV) is approximately $62.3\%$. The F1-score and the Area Under Curve (AUC) are of $0.87$ and $0.759$, respectively. 

\noindent
\textbf{Conclusion:} The proposed methodology makes it possible to predict the distribution of the ODX score for a patient and provides an explanation of the predicted score. The use of the methodology with the pathologist’s expertise on the different histological and immunohistochemical characteristics has a clinical impact to help oncologist in decision-making regarding breast cancer therapy. 

\end{abstract}


\section{Introduction}
The Oncotype DX (ODX) test is a commercially available molecular test for breast cancer assay (Genomic Health) that provides prognostic and predictive breast cancer recurrence information for hormone positive, HER2-negative patients. The ODX test is based on 1A-level evidence and it is included in the main international clinical guidelines recommendations such as those of the American  Society of Clinical Oncology (ASCO \cite{andre2019use}) or the National Comprehensive Cancer Network (NCCN) as well as in the last staging guidelines of AJCC 8th edition \cite{giuliano2017breast}.
The ODX test is the most widely available molecular test used in the world. This assay analyzes 21 genes by RT-qPCR (16 cancer-related genes and 5 housekeeping genes) and aims to predict the risk of recurrence at 10 years by providing a recurrence score ranging from 0 to 100 and to estimate the benefit of adjuvant chemotherapy. Several retrospective and prospective studies have validated this test and its clinical utility. \citet{paik2004multigene} have shown a correlation between ODX score and disease-free survival in patients with ER-positive/HER2-negative, node negative, tamoxifen-treated breast cancer, based on the NSABP B-14 trial. As for the chemotherapy benefits, \citet{paik2006gene} and \citet{albain2010prognostic} have evaluated the test using the studies related to NSABP-B20 and SWOG 8814. The prospective phase III trial TAILORx study \cite{sparano2018adjuvant} has modified the ODX score’s cutoff values (low risk $<$11, intermediate risk 11-25 and high risk $>$25) in order to avoid under-treatments of cancer. To be more precise, in the low group, the risk of recurrence at 5 years is very low ($<10\%$) with hormonal therapy, which confirms the uselessness of adding a chemotherapy \cite{sparano2015prospective}. For the intermediate group, chemotherapy has a benefit only for women younger than 50 years old. For the high-risk group, the chemotherapy is highly recommended. 
Nevertheless, one third of women with  hormone-receptor positive  breast  cancer have a lymph node disease. Thus, the prospective trial RxPONDER trial study analyzes the capacity of the ODX test to predict the benefit of chemotherapy for women with positive lymph node disease \cite{kalinsky202121}. RxPONDER showed that postmenopausal patients with node involvement and an ODX score between 0 and 25 did not benefit from chemotherapy, whereas premenopausal patients with node involvement with 1-3 nodes and ODX scores between 0 and 25 benefited significantly from chemotherapy. 

Despite its proven value, the ODX test is not routinely used due to its high cost. For this reason, less than $20\%$ of patients in Europe have access to the ODX test. Health-related economic study are performed to understand for which patients the assay is the most useful \cite{albanell2016pooled}.  From this economic point of view, many alternative tools have been developed to predict this score. These tools are based on clinico-pathological data such as Magee equations \cite{klein2013prediction, sughayer2018applying} and the IHC4 score \cite{yeo2015clinical}. Indeed, many studies have shown the correlation between the results of the latter tools and the ODX score \cite{flanagan2008histopathologic}. 
Few works used features with machine learning techniques in order to provide an ODX-based methodology to divide the patients into categories corresponding to low or high risk of cancer \cite{kim2019predictive, orucevic2019nomogram,baltres2020prediction, pawloski2022supervised}.

The aim of this paper is to propose a novel methodology to assist physicians in their decision-making.  It is based on random forests for distributional regression as presented in              \citet{Meinshausen2006} and \citet{Atheyetal2019}. This methodology creates links between a new patient and the cohort used for training based on clinico-pathological characteristics. These links can be used particularly to identify the patients of the training cohort that share similarities with the new patient and to predict an estimate of the distribution of the ODX score. This information is available to clinicians to help them better understand the probable clinical evolution of the tumor in order to optimize the treatment. 
 
Moreover, it enables  knowledge capitalization by feedback and analysis of patient history. One of the consequences of this study is to weight the variability of the anatomo-pathological data, so this new methodology can adapt to the specificities of a cohort.

\section{Materials and methods}

\subsection{Dataset description}
The cohort is a retrospective study between 2012 and 2020 with 333 cases that underwent an ODX assay from three hospitals in Bourgogne Franche-Comté: Besançon, Belfort and Dijon. All patients have ER-positive and HER2-negative early breast cancer. Clinical and pathological reports were used to collect the data such as the age at diagnosis, the menopausal status, the treatment, the recurrence, the tumor size, the lymph node status, the histological type, the Nottingham grade, hormone receptors for estrogen (ER) expression, hormone receptors for progesterone (PR) expression, the human epidermal growth factor receptor 2 (HER2) status and the protein p53 and Ki67 proliferation index. Immunohistochemical staining was performed (Ventana Benchmark XT system®, Roche™) on the tumor block of ODX testing with UltraView Universal DAB detection with ER antibody (clone SP1; Roche/Ventana Medical Systems, Tucson, USA), PR antibody (clone 1E2; Roche/Ventana Medical Systems, Tucson, USA), HER2 antibody (clone 4B5; Roche/Ventana Medical Systems, Tucson, USA), Ki67 antibody (clone Mib-1, Dako, Glostrup, Denmark) and p53 antibody (clone DO-7, Dako, Glostrup, Denmark). 
The HER2 immunostaining was interpreted using the 2018 American Society of Clinical Oncology/College of American Pathologists guidelines \cite{wolff2018human}. The Ki67 proliferation index was evaluated by manual counting with counter on at least 200 tumor cells. The protein p53 was assessed by immunohistochemistry. The positive threshold is greater than $10\%$ of the tumor cells' nuclei.
The ODX test was realized by Genomic Health (Redwood City, CA, USA) and analyzed 21 genes by RT-qPCR from paraffin-embedded blocks of tumor tissue. The ODX score was obtained from the clinical reports. The three  ODX categories were the same as the ones defined in the ODX‘s assays using TAILORx and RxPONDER: low risk ($<$ 16), intermediate risk (16-25) and high risk ($>$ 25). 
The institution review board approved this study.

The cohort contains more than 50 features, from which we selected the most critical ones using feature importance in random forest and physicians' assessments. Table \ref{characteristics} describes the tumor characteristics using the  features selected for our study.

\begin{table}
\begin{adjustwidth}{-2in}{-2in}
    \centering
\begin{tabular}{|M{3cm} M{3cm}|M{2cm}|M{2cm}|M{2cm}|M{2cm}|}
        \cline{3-6}
        \multicolumn{2}{M{6cm}|}{} &\multicolumn{4}{M{8cm}|}{Percentage of patient by category}\\
        \cline{3-6}
     \multicolumn{2}{M{6cm}|}{} & $< 16$&	$16-25$ &	$>  25$ & Total \\
    \hline
    Population & & 113 & 138 & 82 &333 \\
    		\hline		
    \multirow{2}{3cm}{\parbox[c]{3cm}{\centering Age}}&	$\leq50$ yr & 14.41 & 10.51 & 5.71 & 30.63 \\
    & $>50$ yr & 19.52 & 30.92 & 18.92 & 69.37 \\
    \hline
    \multirow{3}{3cm}{\parbox[c]{3cm}{\centering Tumor size}} & 	$<$  1 cm & 3.90 & 4.51 & 3.00 & 11.41 \\
    & 1-2 cm  & 15.62 & 22.52 & 15.62 & 53.76 \\
    & $>$ 2 cm	& 14.41 & 14.41 & 6.01 & 34.83\\
    \hline
    \multirow{2}{3cm}{\parbox[c]{3cm}{\centering p53}} &	$\leq 10 \%$   & 18.62 & 23.12 & 12.01 & 53.75\\
    & $> 10 \%$ & 15.32 & 18.32 & 12.61 & 46.25\\

    \hline
    \multirow{3}{3cm}{\parbox[c]{3cm}{\centering SBR grade}}	&1& 5.41 & 3.30 & 0.00 & 8.71  \\
    &2& 21.02 & 24.03 & 10.81 & 55.86 \\
    &3& 7.51 & 14.11 & 13.81 & 35.43\\	
    \hline
    \multirow{3}{3cm}{\parbox[c]{3cm}{\centering Mitotic grade}}	&1& 12.61 & 14.42 & 4.50 & 31.53\\
    &2& 17.12 & 20.12 & 12.01 & 49.25 \\
    &3	& 4.20 & 6.91 & 8.11 & 19.22\\	
    \hline
    \multirow{2}{3cm}{\parbox[c]{3cm}{\centering ER  status}} &	Negative & 0.00 & 0.00 & 0.00 & 0.00\\
    &Positive ($\geq10\%$)	&  33.93 & 41.44 & 24.63 & 100\\
    \hline
    \multirow{2}{3cm}{\parbox[c]{3cm}{\centering PR status}} & Negative & 2.10 & 7.51 & 8.11 & 17.72\\
    &Positive ($\geq10\%$) & 31.83 & 33.93 & 16.52 & 82.28 \\			
    \hline
    \multirow{3}{3cm}{\parbox[c]{3cm}{\centering Ki67-positive cells}} &	$< 10 \%$ & 0.00 & 0.30 & 0.00 & 0.30\\
    & $10-20 \%$ & 16.22 & 15.92 & 4.80 & 36.94\\
    & $> 20 \%$& 17.72 & 25.22 & 19.82 & 62.76 \\	
    \hline
    \multirow{5}{3cm}{\parbox[c]{3cm}{\centering Lymph node status}} & $0$ & 15.02 & 15.52 & 13.81 & 45.35 \\
    & $1$ & 10.81 & 14.11 & 4.20 & 29.13 \\
    & $2$ & 3.61 & 3.30 & 0.90 & 7.81\\
    & $3$ & 1.80 & 2.10 &2.10 & 6.00 \\
    & NA & 2.70 & 5.41 & 3.60 & 11.71 \\
    \hline
\end{tabular} 
\end{adjustwidth}
\caption{ODX score distribution by patient and tumor characteristics.}
\label{characteristics}
\end{table}

\subsection{Distributional Random Forest}
Random Forest \cite{Breiman2001}  is a powerful machine learning algorithm that can be used for prediction in various settings and has been successfully applied in the field of medicine \cite{chen2020rfdcr, fernandez2021random, zare2021robust}. Our goal here is to predict the result of the expensive ODX test based on clinico-pathological features. We propose the use of Distributional Random Forest that provides a predictive distribution for the ODX score based on the clinico-pathological features. We shall expose the methodology for Random Forest and Distributional Random Forest.

Standard regression links the mean of the response variable $Y$ to a set of features $X$ based on observations from a training sample of feature–response pairs, say $(X_i,Y_i)$ for $i=1,\ldots,n$. Random Forest (RF) prediction is an ensemble method that consists of the bootstrap aggregation \cite{Breiman1996} of randomized classification and regression trees (CART, \cite{Breimanetal1984}).
The predictive mean can be written as the average
\begin{equation}\label{eq:tree-average}
\hat Y=\frac{1}{B}\sum_{b=1}^B T_b(X),
\end{equation}
where $T^1(X),\ldots,T^B(X)$ corresponds to the prediction of the different trees built on different bootstrap samples. Each single tree prediction takes the form of an average across a neighborhood of $X$ in the tree, i.e.
\[
T_b(X)=\frac{1}{|R_b(X)|} \sum_{X_i\in R_b(X)} Y_i,
\]
with $R_b(X)$ being the region of the feature space that contains $X$ in the tree $T_b$ and $|R_b(X)|$ the numbers of observations that fall into this region. Consequently, the Random Forest prediction \eqref{eq:tree-average} has the equivalent form
\begin{equation}\label{eq:individual-average}
\hat Y=\sum_{i=1}^n w_i(X) Y_i,
\end{equation}
with the Random Forest weights defined by
\begin{equation}\label{eq:RF-weights}
w_i(X)=\frac{1}{B}\sum_{i=1}^B \frac{\mathds{1}_{\{X_i\in R(X)\}}}{|R_b(X)|},\quad 1\leq i\leq n,
\end{equation}
and these weights are non-negative with sum $1$ (probability weights).

The main idea of Distributional Random Forest (DRF) relies on Equation~\eqref{eq:individual-average}: the prediction $\hat Y$ is the sample mean of the weighted sample $Y_i$ with weights $w_i(X)$ which can  be seen as an approximation of the conditional distribution of $Y$ given $X$. The cumulative distribution function $F(y| X)=\mathbb{P}(Y\leq y|X)$ is thus approximated by
\begin{equation}\label{eq:weighted-cdf}
\hat F(y|X)=\sum_{i=1}^n w_i(X) \mathds{1}_{\{Y_i\leq y\}}.
\end{equation}
This idea was first suggested by Meinshausen \cite{Meinshausen2006} who proposed the construction of quantile regression forest by approximating the conditional quantile of $Y$ given $X$  by the  quantiles of the weighted empirical distribution \eqref{eq:weighted-cdf}.

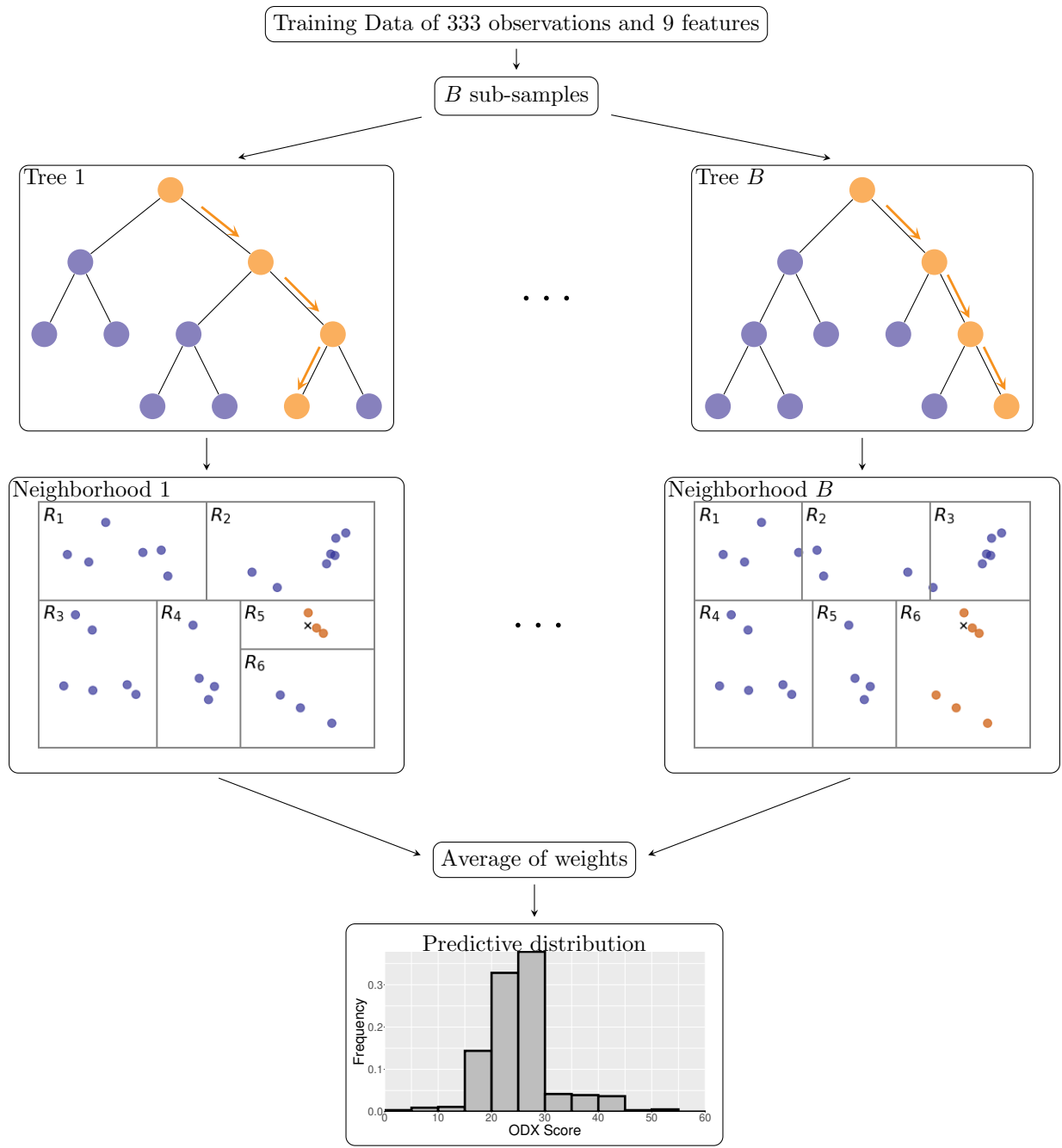
\begin{figure}
    \begin{adjustwidth}{-2in}{-2in}
        \centering
        \begin{forest}
      for tree={l sep=2em, s sep=2em, anchor=center, inner sep=0.4em, fill=refcolor!50, circle, where level=1{no edge}{}}
            [$B$ sub-samples, node box, alias=bagging, above=.5em
            [,pathcolor!70,alias=a1[[,alias=a2][]][,pathcolor!70,edge label={node[above=\arrowdist,path arrow]{}}[[][]][,pathcolor!70,edge label={node[above=\arrowdist,path arrow]{}}[,pathcolor!70,edge label={node[below=\arrowdist,path arrow]{}}][,alias=a3]]]]
            [~~~$\ $~~~,scale=5,no edge,fill=none,yshift=-3em,alias=t2]
            [,pathcolor!70,alias=b1
            [[[,alias=b2][]][]][,pathcolor!70,edge label={node[above=\arrowdist,path arrow]{}}
            [][,pathcolor!70,edge label={node[above=\arrowdist,path arrow]{}}
            [][,pathcolor!70,alias=b3,edge label={node[above=\arrowdist,path arrow]{}}]]]]]
            \node[node box,above=1.5em of bagging.north](data){Training Data of 333 observations and 9 features};
            \node[tree box, fit=(a1)(a2)(a3)] (t1) {};
            \node[tree box, fit=(b1)(b2)(b3)] (tn) {};
            \path (t1) -- (tn) node[midway,scale=2] (t2b) {~~~$\dots$~~~};
            \node[below right=0.2em, inner sep=0pt] at (t1.north west) {Tree $1$};
            \node[below right=0.2em, inner sep=0pt] at (tn.north west) {Tree $B$};
            \draw[black arrow={1mm}{1mm}] (data) -- (bagging);
            \draw[black arrow={5mm}{2mm}] (bagging) -- (t1.north);
            \draw[black arrow={5mm}{2mm}] (bagging) -- (tn.north);
            \node[below=2em of t1.south, node box] (n1) {\includegraphics[width=.42\textwidth]{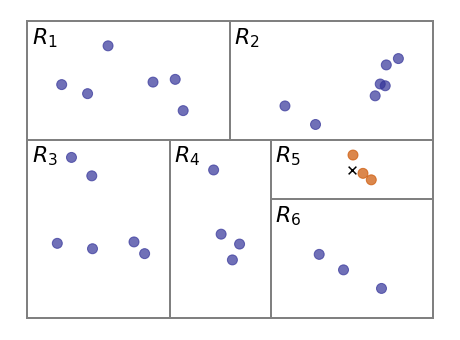}};
            \node[below=2em of tn.south, node box] (n3) {\includegraphics[width=.42\textwidth]{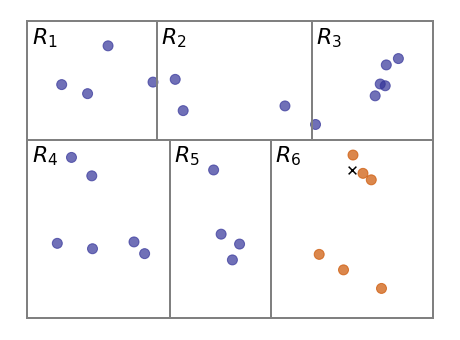}};
            \path (n1) -- (n3) node[midway,scale=2] (n2) {~~~$\dots$~~~};
            \node[below right=0.2em, inner sep=0pt] at (n1.north west) {Neighborhood $1$};
            \node[below right=0.2em, inner sep=0pt] at (n3.north west) {Neighborhood $B$};
            \draw[black arrow={1mm}{1mm}] (t1.south) -- (n1.north);
            \draw[black arrow={1mm}{1mm}] (tn.south) -- (n3.north);
            \path (n1.south west)--(n3.south east) node[midway,below=3em, node box] (mean) {Average of weights};
            \draw[black arrow={2mm}{2mm}] (n1.south) -- (mean.west);
            \draw[black arrow={2mm}{2mm}] (n3.south) -- (mean.east);
            \node[below=2em of mean, node box] (pred) {\includegraphics[width=.4\textwidth]{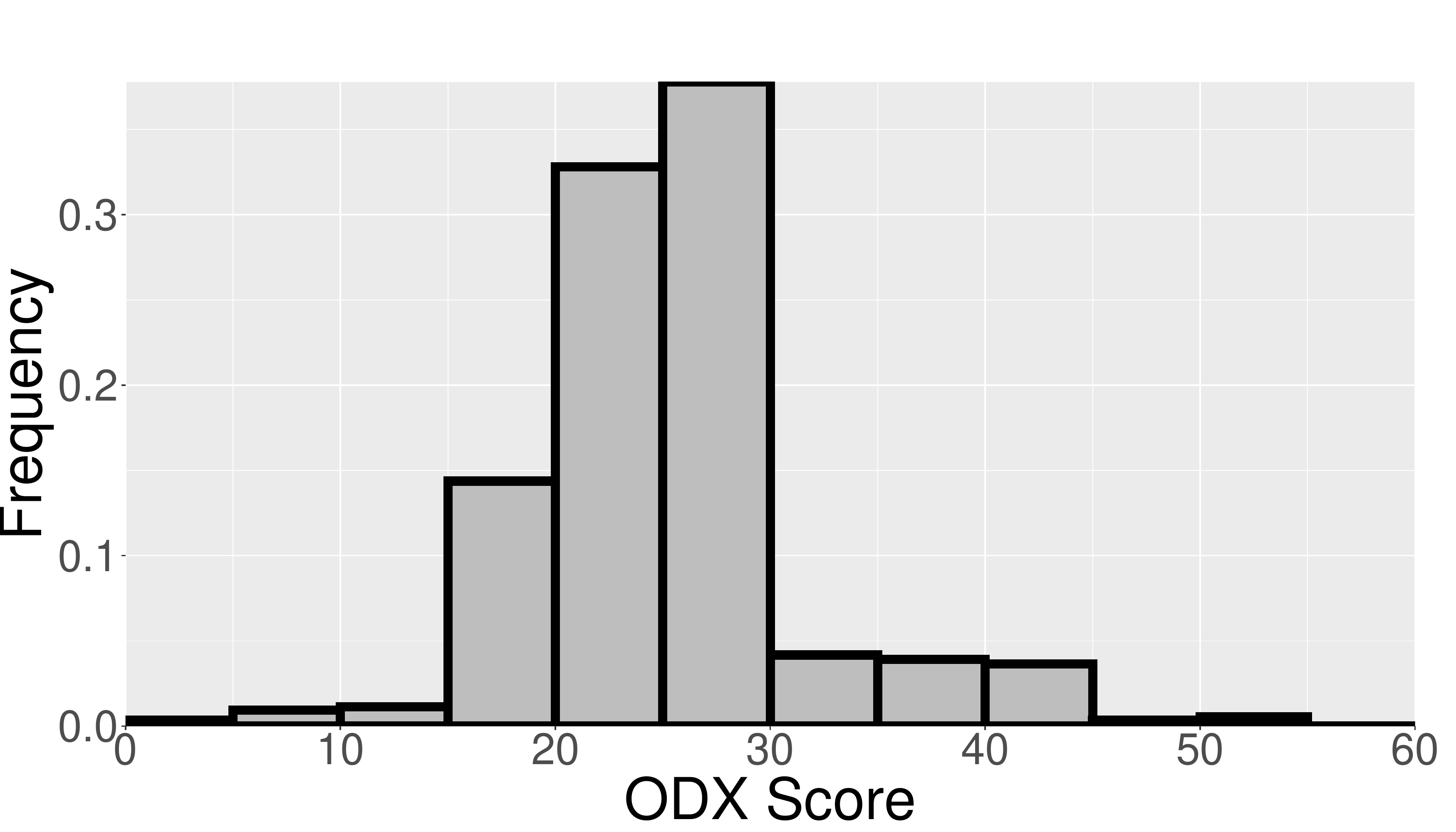}};
            \node[below=0.5em of mean, inner sep=0pt] at (pred.north) {Predictive distribution};
            \draw[black arrow={1mm}{1mm}] (mean) -- (pred);
        \end{forest}
    \end{adjustwidth}
    \caption{Flowchart of Distributional Random Forest. Starting from the training data, a large number of subsamples are randomly chosen and binary trees are constructed on each subsamples; the neighborhood/weights at the point to predict are computed in each tree and then averaged so as to give the forest weights; the predictive distribution corresponds to the weighted sample of the original training data with these forest weights. }
    \label{fig:drf-overview}
\end{figure}

Figure~\ref{fig:drf-overview} presents a synthetic representation of the DRF procedure with the different steps: subsampling of the original sample, tree construction on each subsample, computation of the neighborhood/weight at the point to predict, averaging of weights given by the different trees that finally provide the predictive distribution.

\medskip
The Random Forest weights \eqref{eq:RF-weights} are interesting in themselves and provide relevant information in terms of similar/influential observations. Given a new feature $X$, the weight $w_i(X)$ is interpreted as the proportion in which the observation $Y_i$ contributes to the prediction of $Y$ given $X$. Observations with the largest weights are interpreted as the nearest neighbors of $X$ in terms of an implicit metric on the predictor space that is tailored  for predicting the response, see \cite{LinJeon2006}. The random forest weights make it possible to identify the most similar/influential individuals in the training data. Comparing $X$ to these similar observations can help understand the relationship between $X$ and $Y$.

Finally, let us mention that the weights \eqref{eq:individual-average}-\eqref{eq:RF-weights} depend on the specific structure of the trees that are used for prediction. Trees are grown by recursive binary splitting, maximising a homogeneity criterion; the goal is to partition the feature space into different regions that are as homogeneous as possible. In the standard CART algorithm, the variance is used as the homogeneity criterion, resulting in a partition adapted to the prediction of the mean. Several different splitting rules have been considered in the statistical literature that put the emphasis on the prediction of quantiles \cite[Generalized Random Forest]{Atheyetal2019} or on the overall distribution \citep[Distributional Random Forest]{Cevidetal2021}. 

A Distributional Random Forest is fitted to the whole data set. The software R with the package \texttt{grf} (Generalized Random Forest) is used to compute the random forest and the associated weights. When no new test set is provided, the \texttt{grf::predict} method performs out-of-bag prediction on the training set. This means that, for each training example, all the trees that did not use this example during the training are identified (the example was ‘out-of-bag’), and a prediction for the test example is then made using only these trees.

\subsection{Applications of Distributional Random Forest}
\begin{figure}[H]
    \begin{adjustwidth}{-2in}{-2in}
    \centering
    \begin{tikzpicture}[background rectangle/.style={fill=white}, show background rectangle]
            \node[align=center] (blank_hist) {\includegraphics[width=.5\textwidth]{histogram_blank.pdf}};
			\node[align=center,below=7em of blank_hist.south] (classes_hist) {\includegraphics[width=.4\textwidth]{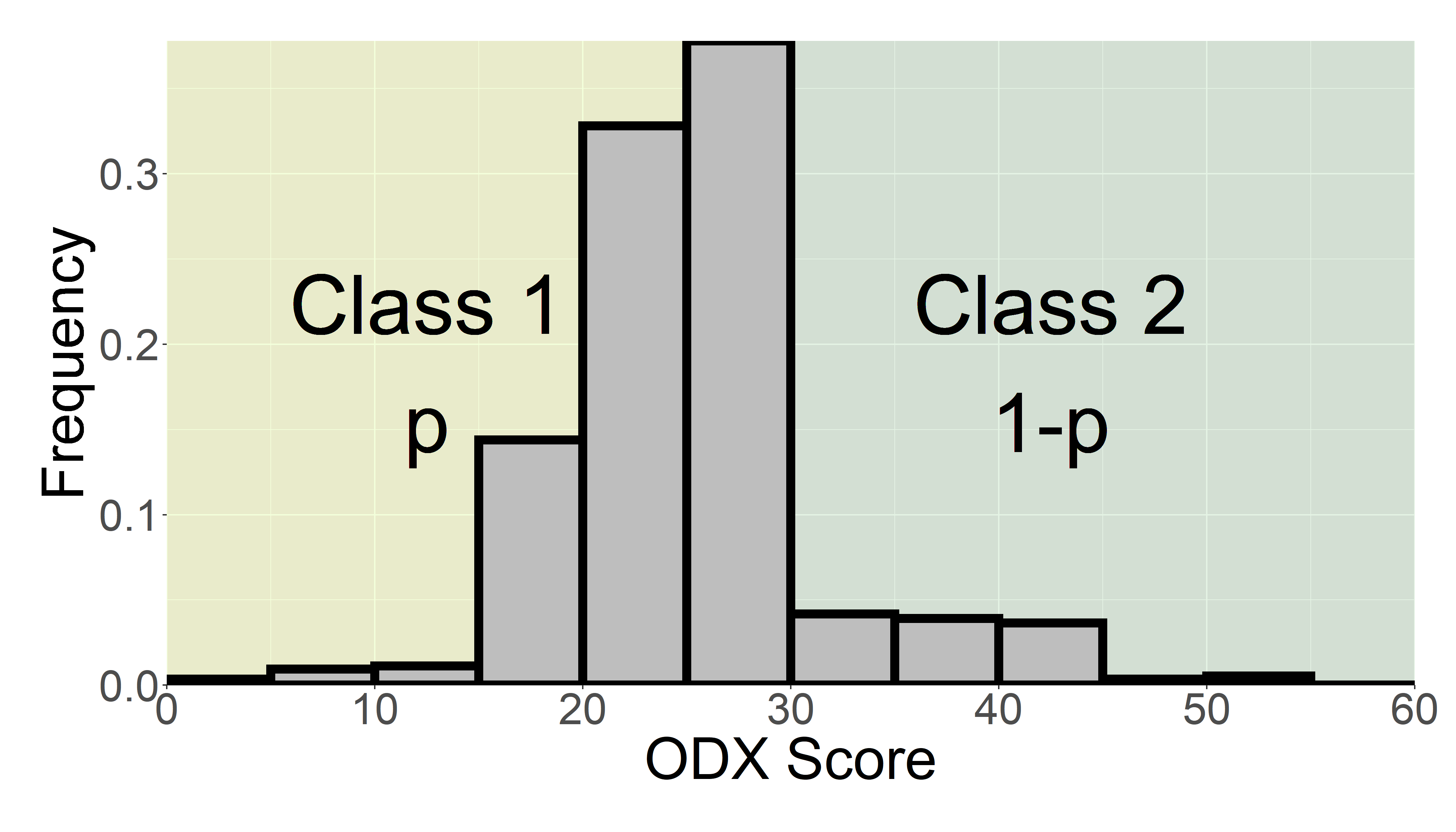}};
			\node[align=center,above right=-2.5em and 0.5em of classes_hist.east] (neighbors) {\footnotesize \renewcommand{\arraystretch}{1.5} \begin{tabular}{|c|c|c|c|c|}
					\hline
					$i$ & $w_i$ & Ki67 & p53 & \dots \\
					\hline
					1& 0.2 & 17 & 8 & \dots \\
					\hline
					2& 0.18 & 20 & 2 & \dots \\
					\hline
					\dots & \dots & \dots & \dots & \dots \\
					\hline
			\end{tabular}};
			\node[align=center,below left=-2.5em and 3em of classes_hist.west] (mean_pred) {\huge ($\hat{Y}$, $\hat \sigma_Y$)};
			
			\node[align=center,above=-0.5em of classes_hist.north, inner sep=0pt] (label1) {\textbf{Classification}};
			\node[above=0.2em of neighbors.north, inner sep=0pt] (label2) {\textbf{Most similar patients}};
			\node[above=0.2em of mean_pred.north, inner sep=0pt] (label3) {\begin{tabular}{c}
			     \textbf{Mean prediction} \\
			     \textbf{Uncertainty assessment}
			\end{tabular}};
            \draw[black arrow={1mm}{1mm}] (blank_hist.south) -- +(0,-1) -|  (label1.north);
            \draw[black arrow={1mm}{1mm}] (blank_hist.south) -- +(0,-1) -| (label2.north);
            \draw[black arrow={1mm}{1mm}] (blank_hist.south) -- +(0,-1) -| (label3.north);
        \end{tikzpicture}
    \end{adjustwidth}
    \caption{\textcolor{red}{}Applications of Distributional Random Forest. Once the DRF is trained, prediction of classes (classification) or of conditional mean or median (regression) together with an uncertainty estimate is straightforward. Furthermore, the weights at the point to predict make it possible to identify  the most similar neighbors in the training data, with an adaptive notion of similarity tailored for the purpose of prediction.}
    \label{fig:drf-appli}
    
\end{figure}

DRF is a fully non-parametric and model-free method that performs probabilistic forecast and distributional regression. For a set of features $X$, it provides the full predictive distribution of the response variable $Y$, that is to say exhaustive information for its possible fluctuations knowing the features. The method is very informative and powerful (see Figure \ref{fig:drf-appli}) as it provides:
\begin{itemize}
    \item (distributional regression) a predictive distribution for each new case that can be represented by a histogram;
    \item (mean or median prediction)  a predictive mean or median when a point estimate is needed - the mean is commonly used while the median is more robust to outliers; 
    \item (uncertainty assessment) a graphical assessment of the uncertainty with the shape of the histogram (either peaked or flat) or numerical statistics  such as standard error or confidence interval for the prediction;
    \item (classification)  the probability of classes of particular interest can be instantly computed - for the ODX score, the classes $\mathrm{ODX}\leq 25$ and $\mathrm{ODX}> 25$ are considered;
    \item (similar/influential patients) the  patients in the cohort (training set) that are the most similar to a new case can be easily identified through the random forest weights that are interpreted as a measure of proximity - this proximity is meant in the sense of an implicit distance that is learnt by the model and that gives more importance to the relevant features; this information can allow the practitioner to make meaningful and informative comparisons between the new case and the patients from the cohort.
\end{itemize}

\subsection{Evaluation of predictive performance}\label{sec:epp}

In order to evaluate the distributional random forest algorithm and compare it with concurrent methods, the theory of a proper scoring rule  \cite{GneitingRaftery2007} is used. In probabilistic forecasting, a scoring rule compares a predictive distribution $F$ and the outcomes $y$. It plays the role of a measure of error similar to  the mean squared error in regression  or the misclassification rate in  classification. A scoring rule is strictly proper if the expected score is minimal when the predictive distribution $F$ matches the outcome distribution. A strictly proper scoring rule can be used for the evaluation of probabilistic forecast and distributional regression \cite{GneitingKatzfuss2014}. 

The most popular scoring rule is the Continuous Ranked Probability Score (CRPS) \cite{Matheson1976} and is defined by 
\[
\mathrm{CRPS}(F,y)=\int_{\mathbb{R}} (F(z)-\mathds{1}_{\{y\leq z\}})^2 \mathrm{d}z.
\]
 In a case where the predictive distribution $F$ corresponds to a weighted sample $(y_i)_{1\leq i  \leq n} $ with weights $(w_i)_{1\leq i\leq n}$, the CRPS is easily computed by 
\[
\mathrm{CRPS}(F,y)=\sum_{i=1}^n w_i|y_i-y| -\sum_{1\leq i<j\leq n}w_iw_j|y_i-y_j|.
\]
The first term compares the predictive distribution $F$ and observation $y$ (calibration) while the second term assesses the precision of the prediction (sharpness). This expression also shows that   $\mathrm{CRPS}(F,y)$ is reported in the same unit as the observation $y$ and that it generalizes the absolute error to which it is reduced if $F$ is a deterministic forecast, that is to say a point measure.

In order to evaluate the generalization capacity of the model, that is to say its predictive performance on a new sample, different validation methods can be used to assess the prediction error. Simple validation uses a training set to fit the model and an independent test set to compute error (CRPS). $K$-fold cross validation is more involved and splits  the data into  $K$ groups that successively play the role of the test set. More precisely, $K$ different models are fitted on training sets consisting of all folds but one  which is left-out during training and used as a test set to compute the CRPS; this results in $K$ different test errors which are averaged so as to obtain the $K$-fold cross validation error.
In the specific case of bagging including our random forest method, the out-of-bag (OOB) method can be used instead. It usually provides similar results as  $K$-fold cross validation but is much more numerically efficient since only one fit of the model is required. Indeed, due to resampling, a given observation does not belong to all the subsamples and one can consider the submodel aggregating all the trees that were trained without this observation; this submodel is then evaluated at the observation and the error (CRPS) is computed; averaging all these errors yield the OOB error.

\section{Results}
The DRF was applied to 333 patients to predict the ODX score  using the 9 features presented in Table \ref{characteristics}. In order to compare with the literature, we emphasize the classification into two classes ($\mathrm{ODX}\leq 25$ and $\mathrm{ODX}>25$). 

\begin{figure}
    \centering
    \makebox[\textwidth][c]{\includegraphics[width=14cm]{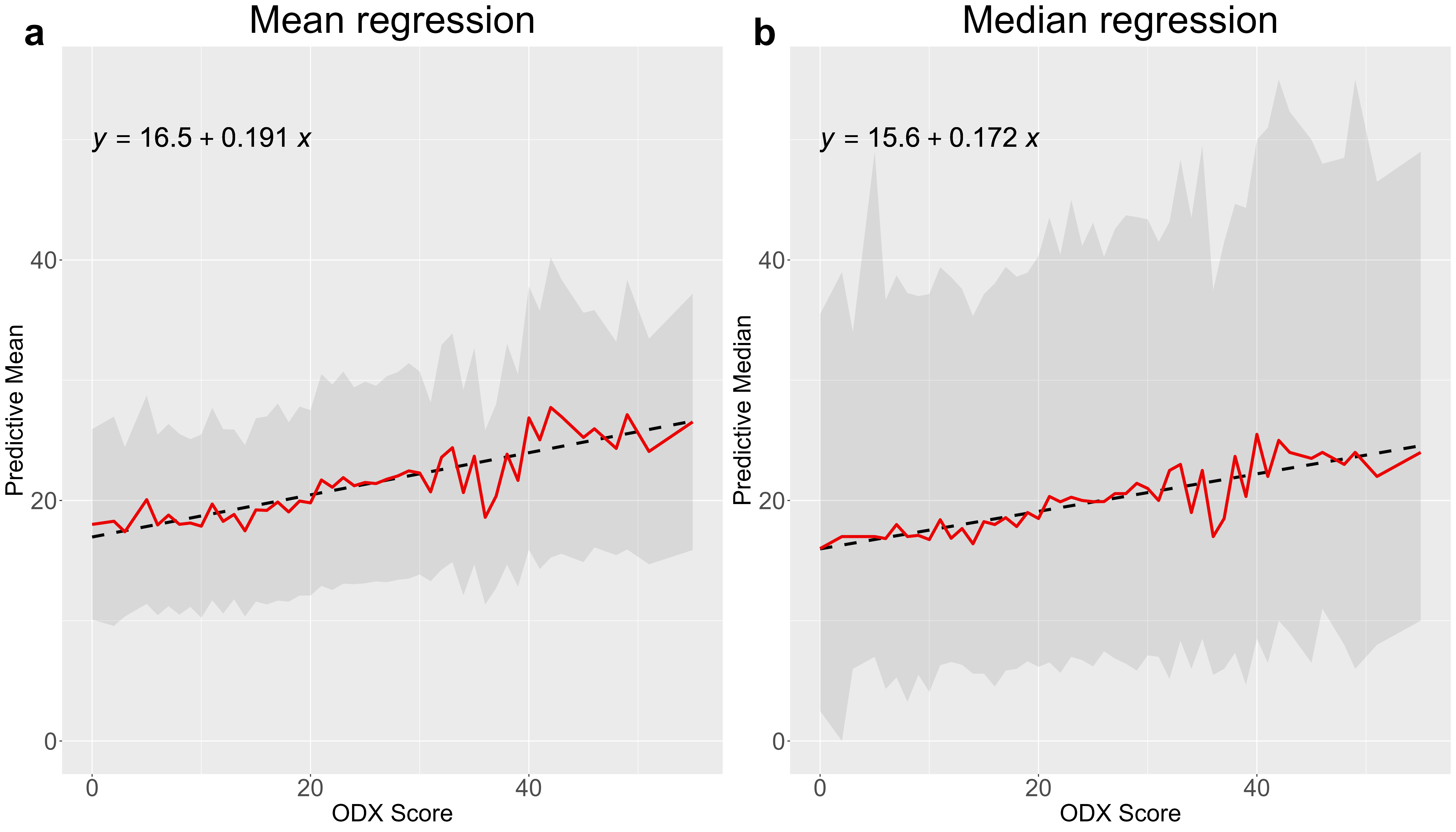}}
    \caption{Evaluation with regression diagnostics.  Mean-regression (left figure - a) plots the ODX observation versus their predictive mean; the grey ribbon represents the standard errors. Median-regression (right figure - b) plots the ODX score versus their predictive median; the grey ribbon represents the $90\%$-confidence interval.}
    \label{fig:evaluation-regression}
\end{figure}

Before presenting the results of the DRF, we shall first present the evaluation of our model. Simple graphical diagnostics can be performed by considering the regression model deduced from DRF. The results of the regression are presented in Figure~\ref{fig:evaluation-regression}, where the predictive mean (Figure~\ref{fig:evaluation-regression}a) and predictive median (Figure~\ref{fig:evaluation-regression}b) versus the real ODX score are plotted. We can observe a rather good fit, and that an important proportion of the observations are within in their confidence intervals. In Figure \ref{fig:evaluation-regression}a,  the grey ribbon has a  semi-amplitude equal to the standard error and accounting for uncertainty. The grey ribbon in Figure \ref{fig:evaluation-regression}b represents the credibility interval with a level of $90\%$. 

\begin{figure}
    \centering
    \makebox[\textwidth][c]{\includegraphics[width=15cm]{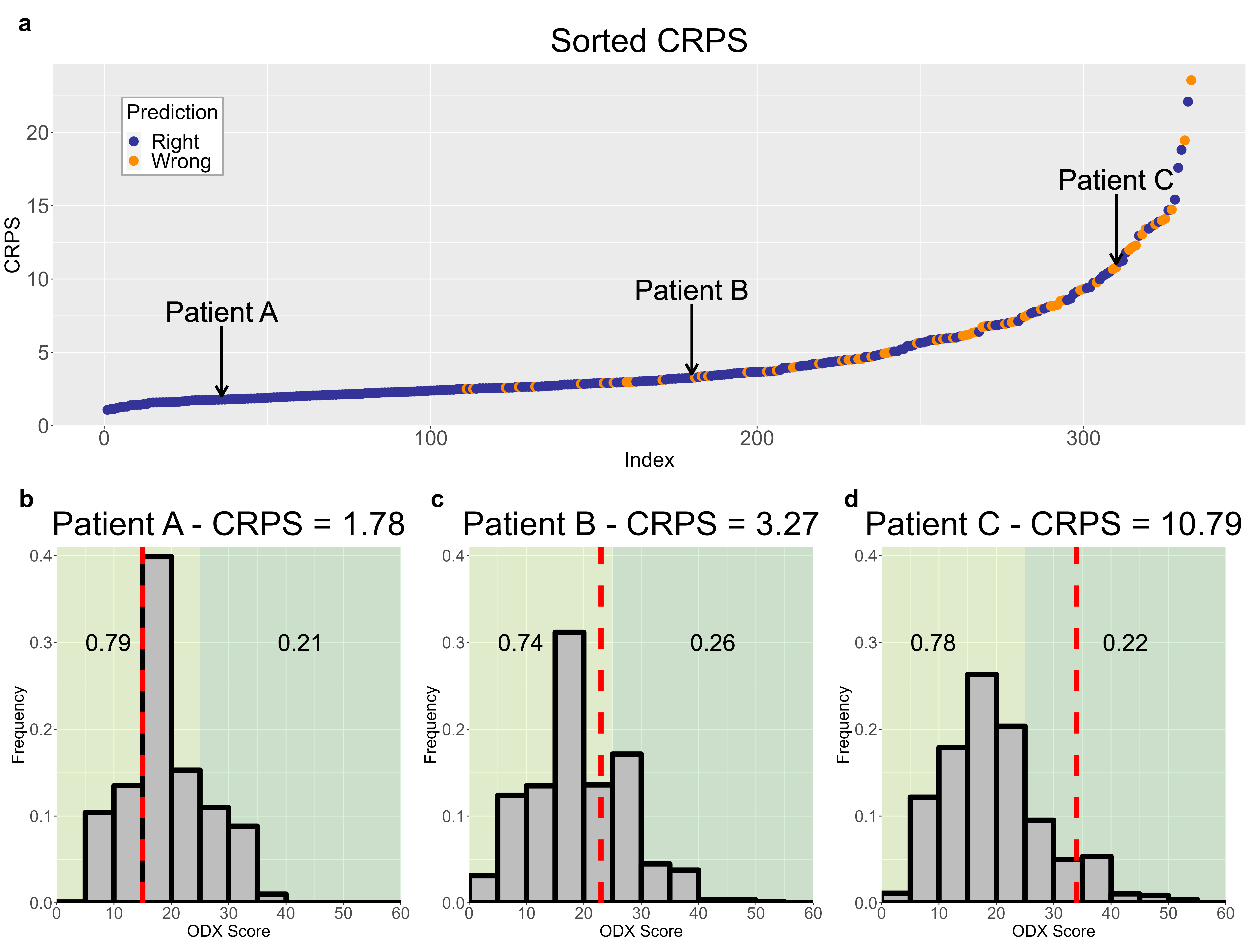}}
    \caption{Out-of-Bag evaluation of the prediction with the CRPS - a low CRPS corresponds to a precise forecast. The subfigures b, c and d correspond to three different patients chosen in different range of the CRPS presented in the subfigure a. In the three lower subfigures, the gray histogram corresponds to the predicted distribution of the ODX score obtained by the DRF. The red dashed line represents the true ODX score of the patient. The two-classes ($\mathrm{ODX}\leq25$ and $\mathrm{ODX}>25$) are represented as areas of different colors and the predicted probabilities of each class is given for each patient.}
    \label{fig:good-avg-bad-crps-oob}
\end{figure}

Additionally, in order to assess the ODX probabilistic forecast, we compared the OOB predictive distribution and the actual observation for the ODX, for each observation. The prediction error is measured in terms of the CRPS introduced in Section~\ref{sec:epp}. The different scores are represented in Figure~\ref{fig:good-avg-bad-crps-oob}a. The smaller the CRPS, the more accurate the forecast. We can observe that most of the predictions have a small or medium CRPS, which indicates the overall good quality of prediction. A smaller number of observations have a large CRPS, indicating individuals for whom the ODX score notably differs from what we might expect in comparison with the overall population. Together with the CRPS, the figure provides the results for the binary classification task ($\mathrm{ODX}\leq 25$ or $\mathrm{ODX}> 25$): classification errors are indicated with the color orange while the color blue corresponds to correctly classified observations. We can observe a good match between classification errors and a large CRPS, which confirms the ability of the CRPS to assess forecast quality. Then, for each patient, the DRF provides a predictive distribution represented by a histogram that can be compared with the actual ODX score. We also indicate the two class probabilities corresponding to the light-green/left or dark-green/right classes. 
We have selected three patients respectively  with a low  (Figure~\ref{fig:good-avg-bad-crps-oob}b, Patient A), medium  (Figure~\ref{fig:good-avg-bad-crps-oob}c, Patient B)  and large CRPS (Figure~\ref{fig:good-avg-bad-crps-oob}d, Patient C). The predictions associated to these patients can be considered "good", "average" and "bad", respectively. In Figure~\ref{fig:good-avg-bad-crps-oob}b we can observe a sharp predictive distribution (peaked histogram) and an ODX score close to the peak. In Figure~\ref{fig:good-avg-bad-crps-oob}c, the histogram is flatter, indicating more uncertainty, and the true ODX score is contained in a high probability region. In Figure~\ref{fig:good-avg-bad-crps-oob}d, the predictive distribution has also a large dispersion and the ODX score is contained in a low probability region, which means that the match between the two is poor. We insist on the fact that a large CRPS does not necessary mean a miss-classification of a patient as it can be seen for some of the higher CRPS values in Figure~\ref{fig:good-avg-bad-crps-oob}a. The CRPS considers the distributional regression and is not explicitly related to the binary classification presented here.\\

Due to the impact of the classification of ODX in the two classes $\mathrm{ODX}\leq 25$ and $\mathrm{ODX}>25$, we shall present the detailed evaluation of the classification model deduced from DRF (see Table \ref{DRFresults}).  This evaluation is based on the standard classification metrics such as confusion matrix and standard metrics. The standard metrics are as follows:
\begin{equation}
    \text{Accuracy} = \frac{TP+TN}{TP +FP + FN + TN},
\end{equation}
\begin{equation}
 \text{Sensitivity} = \frac{TP}{TP +FN},
\end{equation}
\begin{equation}
\text{Specificity} = \frac{TN}{FP + TN},
\end{equation}
\begin{equation}
    \text{Positive Predictive Value} = \frac{TP}{TP +FP},
\end{equation}
\begin{equation}
    \text{Negative Predictive Value} = \frac{TN}{FN+TN},
\end{equation}
\begin{equation}
 \text{ F1-score} = \frac{2*\text{Positive Predictive Value*Sensitivity}}{\text{Positive Predictive Value+Sensitivity}}
\end{equation}
where TP is the number of patients correctly classified as ODX $\leq25$, FP is the number of patients incorrectly classified as ODX $\leq25$, TN is the number of patients correctly classified as ODX $>25$ and FN is the number of patients incorrectly classified as ODX $>25$.\\

\begin{table}
\begin{tabular}{|M{.15cm} M{1.75cm}|M{1.75cm}|M{1.75cm}|}
    \cline{3-4}
    \multicolumn{2}{c|}{}&\multicolumn{2}{c|}{Predicted}   \\
    \multicolumn{2}{c|}{}&$\mathrm{ODX}\leq 25$ &$\mathrm{ODX}> 25$ \\
    \hline
    \multirow{2}{*}{\rotatebox[origin=c]{90}{\parbox[c]{0.75cm}{\centering True}}} &$\mathrm{ODX}\leq 25$ & 231 & 20 \\
    \cline{2-4}
  &$\mathrm{ODX}> 25$ & 49 & 33\\
  \hline
\end{tabular}
    \quad\begin{tabular}{|c|c|}
    \hline
    Accuracy & 79.3\%\\
    Sensitivity & 92.0\%\\
    Specificity & 40.3\%\\
    Positive Predictive Value & 82.5\%\\
    Negative Predictive Value & 62.3\%\\
    F1-score & 0.870\\
    Area Under Curve &  0.759\\
    \hline
    \end{tabular}
\caption{Evaluation with classification diagnostics. Confusion matrix (left) together with standard metrics (right).}
\label{DRFresults}
\end{table}

We have correctly classified 231 out of 251 patients ($92\%$) in low risk and 33 of 75  patients ($40.2\%$) in high risk. The overall accuracy is $79.3\%$ and the p-value is less than $0.05$. The proportion of low risk correct predicted value (PPV) is $82\%$. The percentage of high risk correct predicted value (NPV) is approximately $62.3\%$. The F1-score and the Area Under Curve (AUC) are of $0.87$ and $0.759$, respectively. The DRF will provide additional information such as the nearest neighbor patients, the distribution of the ODX score and the uncertainty prediction (see Figure \ref{fig:drf-appli}). {We now consider the 69 miss-classified patients with low and high risks. First of all, we notice that the majority of these patients have predictions that are close to the decision border (i.e. close to $\mathrm{ODX}=25$). These patients are miss-classified because of the binary decision and additional information available with the DRF method shows either that the patient's ODX score is close to the decision border or that the neighborhood of the patient is not realistic because of limitations of the training cohort. This first part of the miss-classified patients might have a small CRPS as the CRPS accounts for the dispersion of the prediction and its bias. The second part of the miss-classified patients correspond to extreme values of the ODX score within our cohort. The nearest patients provided by the DRF for these miss-classified patients are thus less informative as they are taken within the cohort that is not representative of these outlier patients. In order to give more quantitative results, we compared the mean absolute difference for the ODX score, Ki67 and p53 between the 69 miss-classified patients and the weighted average value of their neighborhoods. The miss-classified patients have a mean absolute difference of ODX score compared to their neighborhood of $9.84$ where the correctly classified patients have an average absolute difference of $6.29$. In terms of Ki67 and p53, the average absolute difference is $24.56\%$ and $5.77\%$ respectively when the average absolute difference for the correctly classified patient is $16.77\%$ for Ki67 and $5.84$\% for the p53 respectively.}

These classification results are then compared with state-of-the art techniques \cite{klein2013prediction, hou2017using, kim2019predictive, orucevic2019nomogram, baltres2020prediction, pawloski2022supervised}. A detailed comparison is given in Table \ref{comparaisonwhole}.

\section{Discussion}
ODX is the most commonly availabe breast genomic test used in early stage ER postive/HER2-negative breast cancer. It makes it possible to define patients who are unlikely to benefit from chemotherapy. The ODX score is based on 6 gene groups. These groups correspond to the markers analysis in pathological reports. Some have compared the ODX score to this immuno-histological data and proved the predictive relationship with the ODX score. Several studies were published using this clinicopathological data to predict the ODX score with different methods (see Table \ref{comparaisonwhole}). The present study was realised to predict the ODX score from a specific regional cohort of 333 patients with clinical and immuno-histological data using Distributional Random Forest. This prediction is associated with a predictive error on the one hand, and the ability to determine the similar patients on the other hand. The proposed DRF model detected $82\%$ of lower risk patients ($\mathrm{ODX}\leq 25$) and $62.3\%$ of high risk patients ($\mathrm{ODX}> 25$). 

A few studies have proposed some prediction tools for the ODX score  \cite{klein2013prediction, hou2017using, kim2019predictive, orucevic2019nomogram, baltres2020prediction, pawloski2022supervised}. Each study is based on the specific categorization of patients according to the original ODX categories and TAILORx (see ODX Prediction Threshold in Table \ref{comparaisonwhole}). The prediction results of the different studies are similar and based on clinico-pathological data. The tumor size, tumor grade and PR are used in all the six selected published studies as well as for our current study. The Ki67 is not used in \cite{orucevic2019nomogram} and \cite{pawloski2022supervised}. In our study, we integrated the p53.  The  threshold  used for the ODX score is different from one study to another. Our DRF model performs as well as the other prediction tools.  The novelty is in providing additional information to the prediction (see Figure \ref{fig:drf-appli}) such as the probability of classes (low and high risk), the similar profiles and the uncertainty prediction.

The correct predicted values are $82.5\%$ and $62.3\%$ for low and high risk, respectively. We used the CRPS score to distinguish the best and worst prediction. The best results were obtained for ODX profiles below 16. The average Ki67, for the first best ten results, is under $14\%$, which corresponds to the low-risk profile of our previous study \cite{baltres2020prediction}. The average percentages of ER and PR are $93\%$ and $77\%$, respectively, which fits into the same low risk profiles. When looking at the surrounding family and the profile of close patients, we observe that similar profiles vary between $0$ and $25$ for ODX. The similarities fall in the low risk profile. The Ki67 scores of similar profiles for the first ten results are below $25\%$.

\begin{landscape}
\begin{table}
    \begin{tabular}{|M{3cm}|M{2.3cm}|M{2cm}|M{2cm}|M{2.5cm}|M{2cm}|M{2cm}|M{2cm}|M{2cm}|}
    \hline
    & & \titlestudy{klein2013prediction} &  \titlestudy{hou2017using} &  \titlestudy{kim2019predictive} &  \titlestudy{orucevic2019nomogram} &  \titlestudy{baltres2020prediction} &  \titlestudy{pawloski2022supervised} & Current study (DRF)\\
    \hline
       Patients & $(n_{train},n_{test})$& (817, 255) & (-, 163) & (208,76)& (65,754, 18,585) & (152, 168)& (2,587, 1,293) & (333, OOB)\\
         \hline
        \multirow{3}{3cm}{\parbox[c]{3cm}{\centering Age}}  & Mean & -- & 58.6 & -- & -- & -- & -- & 56.9\\
        & Median & -- & -- & 44.0 & 58 & 57.5 & 62 & 58.0\\
        & Range & -- & 34-82 & -- & 19-90 & 30-84 & 56-69 & 30-84 \\
         \hline
       \multirow{10}{3cm}{\parbox[c]{3cm}{\centering Clinico-pathologic features used for modelling}}  & Tumor size & \checkmark & \checkmark & \checkmark & \checkmark & \checkmark & \checkmark & \checkmark \\\cline{3-9}
        &Tumor grade &\checkmark & \checkmark&\checkmark &\checkmark  &\checkmark &\checkmark & \checkmark\\\cline{3-9}
        & Lymphovascular invasion & & & \checkmark & & & \checkmark &\\\cline{3-9}
        &Lymph node status & & & \checkmark &  & \checkmark &  & \checkmark\\\cline{3-9}
        &ER& \checkmark & \checkmark &\checkmark& \checkmark & \checkmark && \checkmark \\\cline{3-9}
        &PR& \checkmark & \checkmark & \checkmark & \checkmark & \checkmark & \checkmark & \checkmark \\\cline{3-9}
        &Ki67& \checkmark & \checkmark & \checkmark & &  \checkmark&& \checkmark\\\cline{3-9}
        &p53 & & & &  & && \checkmark
        \\
         \hline
         \multirow{2}{3cm}{\parbox[c]{3cm}{\centering ODX Prediction}} & Type & Continuous & Continuous & Classification & Classification & Classification & Classification & Distributional\\
         \cline{2-9}
            &  Threshold  & \makecell{$<18$\\$18-30$\\ $>30$} & \makecell{$<18$\\$18-30$\\ $>30$} & \makecell{$<11$\\$>25$}& \makecell{$\leq25$\\$>25$} & \makecell{$<18$\\$18-30$\\$>30$} & \makecell{$\leq25$\\$>25$} & \makecell{$\leq25$\\$>25$} \\
         \hline
         
          Method  &  & Multiple Linear Regression
          & Multiple Linear Regression & Neural Network Decision Jungle & Binomial Logistic Regression & Deep Multi-Layer Perceptron & Random Forest & Distributional Random Forest \\\hline
         \hline
    \multirow{2}{3cm}{\parbox[c]{3cm}{\centering Precision}}  & Low risk & 62.5-69.4\% & 72.6\% & 100\% & 87.5\% & 58.3\% & 92.9\% & 82.5\%\\\cline{3-9}
           & High risk & 68.8-77.8\% & -- & 25.0\% & 79.6\% & 63.0\% & 65.1\% & 62.3\% \\
         \hline
    Sensitivity  &  & 58.6-59.1\% & 85.7\%&11.0\% & 99.2\% & 55\% & 96.3\% & 92.0\%\\
         \hline
    Specificity  &  & 70.5-77.4\%&  41.4\%& 100\%& 18.3\%& 78\% & 48.3\% & 40.2\%\\
         \hline
    AUC & & -- & --& 0.744& 0.81 & 0.63 & -- & 0.759\\
         \hline
    
    \end{tabular} 
        \caption{Comparison of our study with six selected published studies \cite{klein2013prediction,hou2017using,kim2019predictive,orucevic2019nomogram,baltres2020prediction,pawloski2022supervised} to predict the ODX score. For three classes only the sensitivity and specificity of the lower class are given.}
        \label{comparaisonwhole}
    \end{table}
\end{landscape}

As for the results that are discordant, they lie in the high risk class. The averages of the ODX score,  Ki67 and PR are respectively $46\%$, $36\%$ and $22\%$. In addition, a negative correlation between ODX and PR for the best and worst results can be observed.  The similar profiles for such cases have a high PR. This behavior is due to the small number of cases in the high risk category.  An example of the worst prediction is a patient with a high ODX score and probability of lying in the high class of $50\%$.  The real ODX score is $49$ and the predicted ODX score is near to the cut-off. The average ODX score for the 10 first similar profiles is 31 and the distribution is centered around 25. The similar profiles are very dispersed, which is difficult to analyze. Most of the nearest neighbors have an SBR grade of 3. The prediction is bad, but nevertheless, the similar profiles have a low ODX score and a high SBR grade.
 The size of the cohort and the training and testing phase could impact the prediction results. In addition, we have an unbalanced cohort in our study, since we have less patients in the high risk class. In that case, the factors of the similar profiles that influences ODX score such as PR, Ki67 and p53 should be considered. The distribution of identical profiles allows the clinician to retrieve similar historical cases in terms of evolution. The proposed model can be applied even when there is missing data. It makes it possible to predict the low risk class with high certitude, which means no chemotherapy to plan. 
Our study is related to the dataset and it is therefore difficult to generalize to a different cohort because of known inter-cohort variability, especially on some biomarkers such as Ki67. 

\section{Conclusion}
This paper proposes a new methodology for oncotype scoring prediction. This methodology is based on distributional random forest and using 9 clinico-pathological features. It makes it possible to predict the distribution of the ODX score for a patient and provides an explanation of the predicted score by computing the probability of belonging to the low or high risk category and identifying the nearest similar profiles. The proposed Distributional Random Forest model detects $82\%$ of lower risk patients ($\leq 25$) and $62.3\%$ of patients with high risk ($> 25$).  However, DRF presents certain limitations.  The use of DRF with the pathologist’s expertise on the different histological and immunohistochemical characteristics has a clinical impact to help oncologist in decision-making regarding breast cancer therapy. 
The medico-economic interest of this strategy is obvious. Additional studies are needed to further validate the DRF method and improve knowledge extraction from pathological data.\\

\bibliographystyle{plainnat}
\bibliography{references}

\appendix

\end{document}